\documentstyle[aps,prb,preprint,tighten,floats]{revtex}
\begin{document}
\preprint{ {\tt cond-mat/9505022} }
\draft

\title{Effective action approach to a bi-layer frustrated antiferromagnet}
\author{  A. V. Dotsenko }
\address{ School of Physics, The University of New South Wales,
             Sydney 2052, Australia}
\date{Internet: {\tt dav@newt.phys.unsw.edu.au} }
\maketitle

\begin{abstract}
A bi-layer quantum frustrated antiferromagnet
  is studied using an effective action approach.
The action derived from the microscopical Hamiltonian has the
  form of the $O(3)$ non-linear sigma model.
It is solved in the mean field approximation with
  the ultraviolet cut off chosen to fit numerical results.
The obtained phase diagram displays a decrease in the critical value of
  interlayer coupling with increase of in-plane frustration.
The critical point for a single-layer frustrated
  antiferromagnet (the $J_1$-$J_2$ model) is estimated to be
  \mbox{ $ J_{2c} = 0.19 J_1$ }.
\end{abstract}


In the past few years, a lot of attention has been given to
  order-disorder transitions in two-dimensional antiferromagnets.
Much of the interest stems from relevance of the problem
  to high-temperature superconductors where such transitions occur.
Although in superconductors the transition is driven by mobile holes,
  it can also be studied in a frustrated antiferromagnet.
The antiferromagnet
  with frustrating interaction between next nearest neighbours
  (the $J_1$-$J_2$ model)
 is the simplest case and has often been studied.\cite{ds}

In this paper, I study a {\em bi-layer} frustrated antiferromagnet.
The bi-layer problem has emerged from the studies of
 YBa$_2$Cu$_3$O$_{6+x}$ which consists of pairs of close CuO$_2$ planes.
Experimentally,
 there are a number of differences between bi-layer and
 single layer compounds.\cite{experim}
Recent infrared measurements\cite{IR}
 suggest that the antiferromagnet interlayer coupling
 within bi-layers may be as strong as 60 meV
 (about half of the in-plane coupling)
 in which case it must be taken into account in any theoretical study.
Several authors have discussed pairing in bi-layer models.\cite{pairing}
This work is restricted to studying magnetic order in the static
  antiferromagnet.

The Hamiltonian is taken in the form
\begin{eqnarray} \label{H}
  H = J_1 \sum_{\rm NN}
(  {\bf S}_{1{\bf r}} \cdot {\bf S}_{1{\bf r} '}
+  {\bf S}_{2{\bf r}} \cdot {\bf S}_{2{\bf r} '} )
    + J_2 \sum_{\rm 2N}
(  {\bf S}_{1{\bf r}} \cdot {\bf S}_{1{\bf r} '}
+  {\bf S}_{2{\bf r}} \cdot {\bf S}_{2{\bf r} '} ) && \\
    + J_3 \sum_{\rm 3N}
(  {\bf S}_{1{\bf r}} \cdot {\bf S}_{1{\bf r} '}
+  {\bf S}_{2{\bf r}} \cdot {\bf S}_{2{\bf r} '} )
+  J_\perp \sum_{{\bf r}} {\bf S}_{1{\bf r}} \cdot {\bf S}_{2{\bf r}},
   && \nonumber
\end{eqnarray}
where the summation is performed over a square lattice,
 NN denotes nearest neighbours,
 2N (3N) stands for second (third) neighbours, and
 the two planes within the bi-layer are labelled as 1 and 2
 (it is assumed that all interactions $J$ are positive).

In case $J_\perp = 0$ the Hamiltonian (\ref{H}) reduces to a frustrated
  Heisenberg antiferromagnet.
Increasing $J_2$ and $J_3$ leads to a quantum phase transition
  into a disordered phase.
The numerical value of the critical coupling
  $J_{2c}$ (at $J_3 = 0$) is, however, known very poorly.

The case of $ J_2 = J_3 = 0 $ (a bi-layer antiferromagnet)
 has also received a lot of
 attention.\cite{mfswt,sw,cui,series,mm,ss,scs,cm}
Availability of reliable numerical results\cite{series,ss} for this case
 makes it a convenient model\cite{scs} for testing theoretical results.
The bi-layer antiferromagnet
 has also been used\cite{mm} to explain some experimental
 results in bi-layer superconductors.
However, the value of interlayer coupling required for this
 ($J_\perp \gtrsim 2.5 J_1$) is too large to be realistic.
Apparently, both in-plane and inter-plane effects are important.
In the presence of in-plane frustration the critical value
 of interlayer coupling may be drastically reduced.

In both the bi-layer antiferromagnet
 and single layer frustrated antiferromagnet,
 the mean field spin wave and mean field Schwinger boson
 theories significantly overestimate the region of stability of the
 N\'{e}el ordered phase.
The main reason for this appears to be that these theories assume long range
 order from the start and do not include some types of fluctuations
 (see discussion of longitudinal spin fluctuations in bi-layer
 antiferromagnet in Ref.~\onlinecite{cm}).

In this study, I use an effective action approach which
 is based on mapping the microscopic Hamiltonian onto the
 quantum $O(3)$ non-linear sigma model and which has been
 extensively applied to two-dimensional antiferromagnets.
Using a $1/N$ expansion of the non-linear sigma model,
 Chubukov, Sachdev, and Jinwu Ye\cite{csj} have
 recently presented a comprehensive analysis of the general properties
 of clean two-dimensional antiferromagnets
 in the vicinity of the order-disorder transition.
Good agreement with numerical and experimental data has been obtained.

In this work, I first derive the effective action for the bi-layer
  frustrated antiferromagnet.
It is observed then that for the purposes of the study,
  the action can be reduced to that of the standard one-band non-linear
  sigma model.
The phase diagram is then found using
  the mean field (saddle point) solution with the ultraviolet cut off
  (or, equivalently, the critical coupling)
  adjusted so as to fit numerical results for the bi-layer antiferromagnet.

There are several ways to obtain the action.
I follow the derivation based on coherent state representation.\cite{trieste}
Each spin is represented by a single unit vector ${\bf N}$.
Then all spin configurations are expressed as a combination of four fields
  (two fields per layer)
\begin{eqnarray} \label{decomp}
 {\bf N}_{1{\bf r}} = \eta_{\bf r}
        {\bf n}_{1{\bf r}} (1- a^2 {\bf L}^2_{1{\bf r}} )^{1/2}
             + a {\bf L}_{1{\bf r}},\nonumber\\
 {\bf N}_{2{\bf r}} =- \eta_{\bf r}
        {\bf n}_{2{\bf r}} (1- a^2 {\bf L}^2_{2{\bf r}} )^{1/2}
             + a {\bf L}_{2{\bf r}},
\end{eqnarray}
where
\[
  {\bf n}_1^2 = {\bf n}_2^2 = 1,~~~
  {\bf n}_1\cdot{\bf L}_1 = {\bf n}_2\cdot{\bf L}_2 = 0,~~~
  a^2 {\bf L}_{1,2}^2 \ll 1,
\]
$\eta_{\bf r} = \pm 1$ for the two sublattices, and $a$
   is the lattice constant.
The unit length fields $ {\bf n}_{i{\bf r}} $ in the classical
  limit describe
  orientation of the local N\'{e}el ordering and
  the fields ${\bf L}_{i{\bf r}}$
  represent small local fluctuations.

Now the Hamiltonian is expressed using Eq.~(\ref{decomp})
 and the continuum limit is taken by making an expansion
 in gradients of ${\bf n}_1$ and ${\bf n}_2$ and powers of
  ${\bf L}_1$ and ${\bf L}_2$
\begin{eqnarray} \label{H-cont}
 H = \int d^2 {\bf r}
   \biggl[
     \case{1}{2} \rho_{s1} [ (\nabla {\bf n}_1)^2 + (\nabla {\bf n}_2)^2 ]
         + S^2 J_\perp ({\bf n}_1 - {\bf n}_2)^2
    & &  \nonumber \\
         + 4 S^2 (J_1 + \case{1}{8} J_\perp ) ({\bf L}_1^2 + {\bf L}_2^2)
         + S^2 J_\perp {\bf L}_1 \cdot {\bf L}_2
   \biggr],    & &
\end{eqnarray}
where $\rho_{s1} = S^2 (J_1 - 2J_2 - 4J_3)$ is the bare spin stiffness.

Decomposition of the partition function
  is a straightforward extension of the procedure for the single
  layer antiferromagnet\cite{trieste} and leads to
\begin{equation} \label{Z1}
 Z = \int {\cal D} {\bf n}_1 {\cal D} {\bf n}_2
          {\cal D} {\bf L}_1 {\cal D} {\bf L}_2
          \delta ({\bf n}_1^2 - 1) \delta ({\bf n}_2^2 - 1) \exp (S_n)
\end{equation}
with the action
\[
S_n = S_B -
          \int d^2 {\bf r}
          \int_0^\beta  d\tau
  \bigl[ H (\tau) + i S {\bf L}_1
          \cdot ({\bf n}_1 \times \partial_\tau {\bf n}_1)
                  + i S {\bf L}_2
           \cdot ({\bf n}_2 \times \partial_\tau {\bf n}_2)
  \bigr],
\]
  where $S_B$ is the residual Berry phase which is zero for all smooth
  spin configurations and which is ignored hereafter.
Finally, the fields ${\bf L}_1$ and ${\bf L}_2$
  are integrated out of Eq.~(\ref{Z1})
\begin{eqnarray} \label{Z1-5}
 Z = \int&& {\cal D} {\bf n}_1 {\cal D} {\bf n}_2
   \delta ({\bf n}_1^2 - 1) \delta ({\bf n}_2^2 - 1)   \nonumber\\
   &\times& \exp
 \Biggl\{
           -  \int d^2 {\bf r} \int_0^{c\beta} d \tau
  \biggl[
   \case{1}{2} \rho_{s1} [(\nabla {\bf n}_1)^2 + (\nabla {\bf n}_2)^2]
         + S^2 J_\perp ({\bf n}_1 - {\bf n}_2)^2
  \nonumber \\
       &+& \case{1}{8} (J_1 + \case{1}{4} J_\perp)^{-1}
            ({\bf n}_1 \times \partial {\bf n}_1
               + {\bf n}_2 \times \partial {\bf n}_2)^2
         + \case{1}{8} J_1^{-1}
            ({\bf n}_1 \times \partial {\bf n}_1
                - {\bf n}_2 \times \partial {\bf n}_2)^2
  \biggr]
 \Biggr\}.
\end{eqnarray}
At $J_\perp = 0$, the action (\ref{Z1-5})
  represents two independent sigma models
  each possessing a Goldstone mode at $T=0$ due to spontaneous
  symmetry breaking.
At finite $J_\perp$, one mode acquires a gap
   $\Delta \propto \sqrt{J_\perp J_1}$.
There is no general solution in this case because of
  the presence of the term $  ({\bf n}_1 \times \partial {\bf n}_1 )
   \cdot ({\bf n}_2 \times \partial {\bf n}_2)^2) $ in the action.
However, at larger $J_\perp$ and $T\ll\sqrt{J_\perp J_1}$
   the action can be simplified due
  to the fact that it is dominated by configurations with
  ${\bf n}_1 \approx {\bf n}_2$.
In this regime of coupled planes,
  I restrict myself to the Goldstone mode and set
  ${\bf n} = {\bf n}_1 = {\bf n}_2$
  in Eq.~(\ref{Z1-5}).
The action takes the form
\begin{equation} \label{Z2}
 Z = \int {\cal D} {\bf n} \delta ({\bf n}^2 - 1) \exp
 \left\{
           - {1\over 2g} \int d^2 {\bf r} \int_0^{c\beta} d \tau
        \left[
               (\nabla {\bf n})^2 + (\partial_\tau {\bf n})^2
        \right]
 \right\} ,
\end{equation}
where time has been rescaled ($c \tau \rightarrow \tau$ )
  and the coupling constant is   $ g = c/\rho_s $
 with the spin wave velocity
\begin{equation} \label{velocity}
 c = 2 S \bigl[ 2 (J_1-2J_2-4J_3) (J_1 + \case{1}{4} J_\perp) \bigr]^{1/2}
\end{equation}
  and the bare spin stiffness
\begin{equation} \label{rho}
  \rho_s = 2 S^2 (J_1 - 2J_2 - 4J_3).
\end{equation}
{} From Eq.~(\ref{velocity}) for the bi-layer antiferromagnet
  the spin wave velocity at the critical point $J_\perp = 2.5 J_1$
  is $c_c = 1.8J_1$.
It is in good agreement with numerical results\cite{ss} which
  suggests that $1/S$ corrections are negligible at this point.
The spin stiffness Eq.~(\ref{rho})
  is twice as large as in the regime of independent planes.
Thus when $J_\perp$ is introduced
  the system first becomes more ``classical''
  (in other words, the effective spin $S^\ast$ increases)
  and then evolves towards the quantum disordered phase.
This effect was observed in earlier studies\cite{cui,mm,cm}
  at a fairly small value of $J_\perp \approx 0.1 J_1$.
With in-plane frustration present the crossover will occur
  at even smaller values of $J_\perp$.
(In a recent exact diagonalisation study of
   the bi-layer $t$-$J$ model,\cite{eder}
  it was noticed that the two layers become essentially correlated at
  $J_\perp \approx 0.2 J$.)

Equation (\ref{Z2}) represents the quantum $O(3)$ non-linear sigma model.
Not being interested in details of the critical behaviour,
 I use the simplest mean field solution which is exact for
 the $O(\infty)$ non-linear sigma model.
An ultraviolet cut off must be introduced in momentum integration
  and a Pauli-Villars cut off is a convenient choice.\cite{trieste}
In the mean field approximation,
  there are spin-1 excitations with a gapped spectrum and no damping.
Straightforward calculations show that the gap $m$ is determined form
\begin{equation} \label{m}
  \log \left( \sinh {m \over 2T} \right) = - { 2\pi \over T}
    \left(
        {1\over g} - {1\over g_c}
    \right)
\end{equation}
where $g _c = 4\pi/ \Lambda$ is the critical
 coupling ($\Lambda$ is the cut off).
The system is ordered at $T=0$ when $ g<g_c$.
Knowing the critical point at $J_2 = J_3 = 0$ allows
  to eliminate uncertainty
 from the cut off.
Using $J_{\perp c} = 2.5 J$, the cut off is found to be
 $\Lambda = 1.1 \pi$.
It is easy to see that the critical line $g = g_c$ has the form
\begin{equation} \label{crit-line}
  J - 2J_2 - 4J_3 =
{
   J  + \case{1}{4} J_\perp
 \over
   J +  \case{1}{4} J_{\perp c}
}.
\end{equation}
While Eq.~(\ref{m}) is only the mean field solution
  and is not quite accurate,
  Eq.~(\ref{crit-line}) is independent of $1/N$ corrections.
The phase diagram is presented in Figs.~2 and 3.
Figure 2 is the phase diagram for $J_3 = 0$.
In the absence of interlayer coupling,
 the critical value of $J_2$ is found to be
 $J_{2c}=0.19 J_1$.
It is compatible with the results obtained by other methods except for
 numerical results (see Table I).
However, numerical studies\cite{numericalJ1J2}
 face severe size restrictions
 and are not very reliable quantitatively.
Figure 3 is the phase diagram for $J_2 = 0$.
Both Figs. 2 and 3 show that in the presence of an interlayer
 coupling smaller values of in-plane frustration are sufficient
 for the disordering transition.

In conclusion, I have studied the bi-layer frustrated antiferromagnet
 using the effective action approach.
It was demonstrated that the essential physics can be described by
  a one-band non-linear sigma model.
The region of stability of the N\'{e}el ordered phase has been identified
The critical value of next nearest neighbour interaction for
  the $J_1$-$J_2$ model is estimated to be $J_2 = 0.19 J_1$.

\section*{ACKNOWLEDGEMENTS}

I am grateful to O.~P. Sushkov for discussions and
 to A.~V. Chubukov for his comments on the paper.
This work forms part of a project supported by a grant of the Australian
 Research Council.


\begin{table}[tb]
\caption{The critical coupling in the $J_1$-$J_2$ model
    obtained by different methods.}
\begin{tabular}{lc}
method &  $J_{2c}$ \\
\tableline
Green functions\tablenote{Ref.\ \onlinecite{barabanov}}
       &   0.12 \\
renormalization group\tablenote{Ref.\ \onlinecite{swedes2}}
       &   0.15 \\
mean field bond operators\tablenote{Ref.\ \onlinecite{bond}}
       & 0.19 \\
series expansion\tablenote{Ref.\ \onlinecite{seriesJ1J2}}
       & $0.33 \pm 0.08$ \\
present work
       & 0.19 \\
exact diagonalisations\tablenote{Ref.\ \onlinecite{numericalJ1J2}}
       & $ > 0.34 $ \\
\end{tabular}
\end{table}



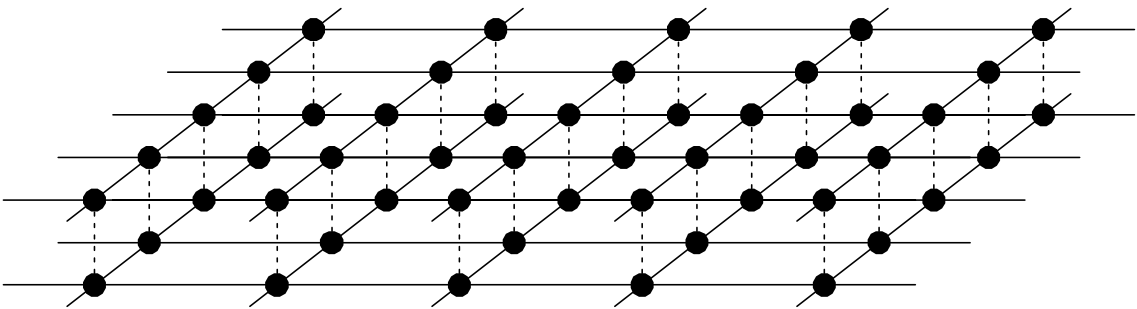
\begin{figure}[h] \caption{  \label{fig1}
 A scheme of the bi-layer antiferromagnet.
}  \end{figure}

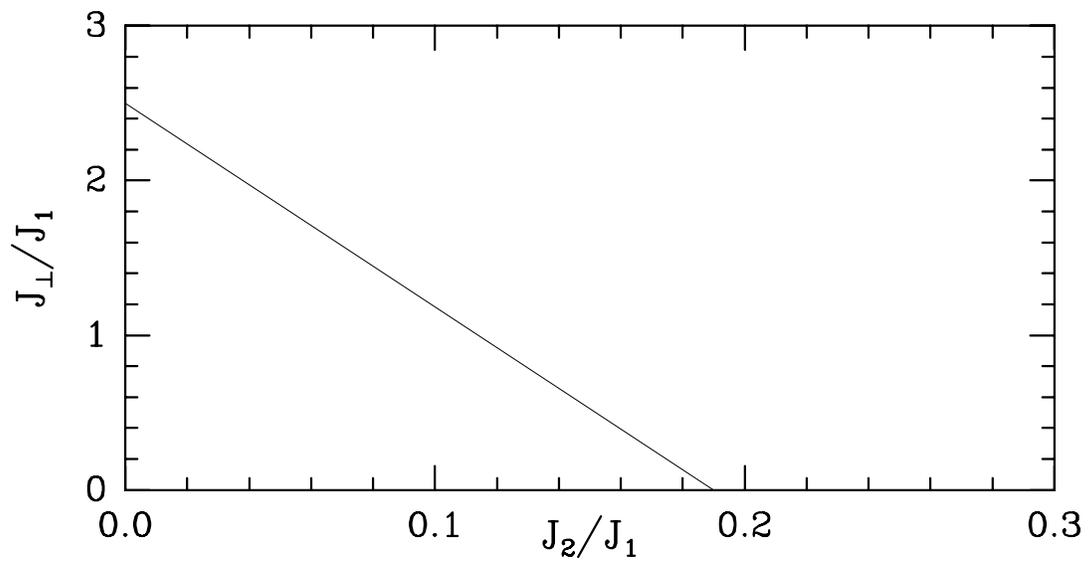
\begin{figure}[h] \caption{  \label{fig2}
 The phase diagram for $J_3 = 0$.
}  \end{figure}

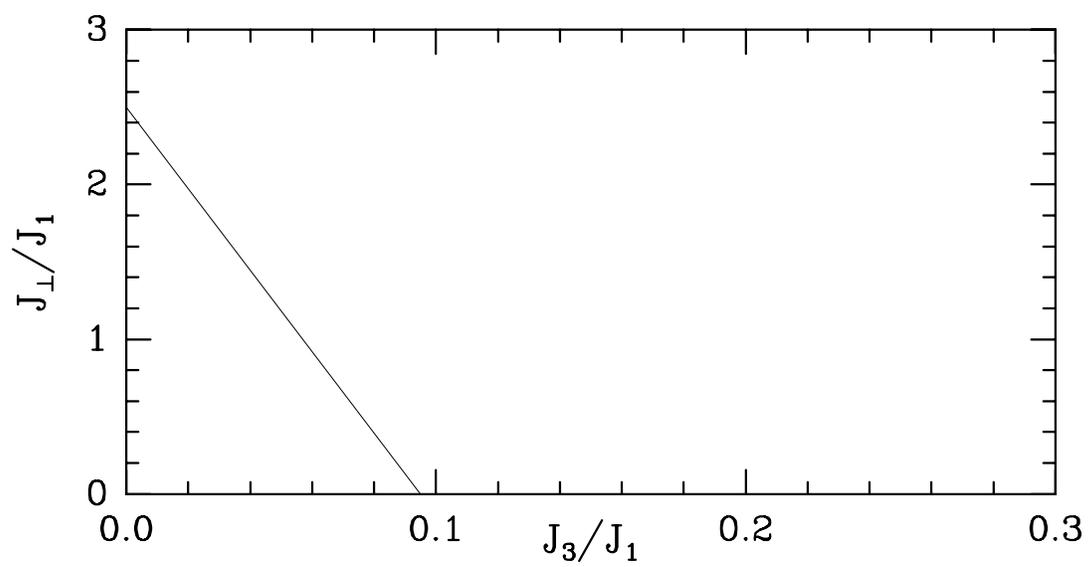
\begin{figure}[h] \caption{  \label{fig3}
 The phase diagram for $J_2=0$.
}  \end{figure}

\end{document}